# Towards a Taxonomy of Sustainability Requirements for Software Design

Mandira Roy
Ca' Foscari University
Venice, Italy

Novarun Deb
University of Calgary
Calgary, Canada

Nabendu Chaki
University of Calcutta
Calcutta, India

Agostino Cortesi
Ca' Foscari University
Venice, Italy

## 1 Introduction

Software inherently contributes to global sustainability concerns [14]. Its impact spans the entire life cycle, beginning with requirements engineering and extending through deployment and usage. The sustainability footprint of a software system largely depends on its features and design choices, making sustainable software design essential to minimize negative effects. There are two main directions for achieving this [14]. The first emphasizes adherence to bench-marked practices and guidelines, such as the Karlskrona Manifesto for green software engineering [5]. The second involves a deeper analysis of the specific software system under development to identify and address sustainability concerns. These two approaches can complement each other. The focus of this research lies on the second direction.

Existing literature provides substantial evidence [4, 6, 7, 9] that sustainability concerns need to be addressed from the requirements engineering phase. Although organizations, realize this need, but lacks concrete support and knowledge. The question that arises- *What are the sustainability requirements (SRs) for a system to be build?* Many research contributions [21], [7], [16] has already been made, formulating sustainability requirements that are necessary. These contribution are often specific to certain sustainability dimension [1], [11], [22] or to application domains [2, 10].

Existing state-of-the-art can form a large database sustainability requirements across various domains. However, there is a lack of systematization of these requirements that could serve as a reference for the software engineering community. This research work extracts and organizes these data (sustainability requirements) into a comprehensive taxonomy. The taxonomy is designed to assist both software developers and researchers in formulating sustainable software development approaches. The main contributions of this work are as follows:

(1) A systematic literature review is conducted to create a taxonomy of sustainability requirements.
(2) A comprehensive set of categories for the four dimensions of sustainability (environmental, technical, social and economic) are identified. These categories are defined along with their metrics and measures.
(3) A correlation matrix among different categories across different sustainability dimensions is depicted. This matrix projects how different sustainability issues may have positive or negative influence on each other.

The structure of the rest of the paper is as follows. Section 2 elaborates on existing state-of-the-art. Section 3 illustrates the methodology for construction of the taxonomy. Section 4 describes the taxonomy created and categories identified. Section 4 describes how the correlation among sustainability categories are defined. Section 5 concludes.

## 2 State of the Art

This section illustrates recent research proposals made for integrating sustainability requirements within software development process.

ShapeRE [19] is a theoretical framework designed for iterative identification and representation of sustainability requirements. Although the work, does not present any actual sustainability requirements but lays a foundation for representing developer-oriented SRs. Researchers [18] have assessed how Linguistic-Relativity-Theory impacts identification of SRs. Authors have experimented with 101 participants, considering the requirements of a grocery application. Their findings show that associating dimensions guides in correctly identifying SRs. Researchers [20] have proposed a goal-based scenario and feature modeling approach for decomposing sustainability requirements. This approach enables the assessment of sustainability risk analysis. An empirical study [3] investigates sociocultural sustainability factors for e learning system. These results provide gender based social sustainability concerns recorded from 177 participants. A study conducted by the authors in [15] has identified key dimensions of sustainability that are being considered within the software engineering process. Their findings show that the characteristics perceived to be the most significant were security, usability, reliability, maintainability, extensibility and portability. But concerns like computer ethics (e.g. privacy and trust), functionality, efficiency, and re-usability have little attention.

A survey conducted in [8] has captured the correlations between software quality requirements and sustainability requirements. The authors have provided a comprehensive list of non-functional requirements (NFRs) category and their association with four sustainability dimensions (environmental, economic, social, technical). Researchers in [17] have listed out sustainability requirements for healthcare applications in smart domains. It covers five dimensions of SRs- social, economic, individual, environmental and technical.

The current state-of-the-art in sustainable software engineering has many directions. Some frameworks or methodologies aim to integrate sustainability concerns within the development process. Some works particularly focused on eliciting SRs for specific domains. There are also survey research where understanding of sustainability concepts among software practitioners are identified. The SRs identified are often only specific to certain domains or framed at a higher abstraction.

## 3 Methodology

This section reports the methodology adopted for building a sustainability requirements taxonomy.

Figure 1 illustrates the process undertaken to identify relevant research works for developing the taxonomy through a systematic literature review (SLR). The subsequent sections provide a detailed explanation of each step.

### 3.1 Research Questions

The formulation of a set of well-defined research questions forms the basis of a focused and effective SLR. These research questions are refined iteratively based on preliminary readings and understanding of the existing state of the art.

RQ-1  How are sustainability requirements defined and characterized across different domains (e.g., stand-alone software, cloud computing, smart systems, healthcare)?

RQ-2  What are the most frequently identified categories of sustainability requirements for different dimensions in the existing state of the art?

RQ-3  What commonalities and divergences of SRs across different domains can be identified and recorded within a generic comprehensive taxonomy?

*RQ-1* is intended towards understanding how sustainability requirements are defined for software systems in general and also considering different possible application domains. *RQ-2* is intended for identifying how much the different dimensions of SRs are explored for designing software systems and what categories within each dimension are considered. *RQ-3* requires comparing SRs defined for different domains, understanding the overlapping among them and also if they are too specific for a particular domain. This step ensures the taxonomy does not contain redundancies in the captured list of SRs. The search strategy employed for answering these research questions directly influences the granularity and comprehensiveness of the taxonomy.

### 3.2 Review Process

The details of the review process are as follows. **Data Sources**
This step defines the databases selected for this study. The authors have explored the most popular digital libraries that consist of peer-reviewed research works for engineering, information Systems, and software applications. The study incorporates only qualitative data sources from the literature. Qualitative data were gathered from peer-reviewed journals, conference proceedings, and book chapters to identify SRs. The specific databases that are searched are:

- Scopus, www.scopus.com
- ACM Digital Library, https://dl.acm.org
- IEEE Xplore, www.ieeexplore.ieee.org
- Wiley Online Library, https://onlinelibrary.wiley.com

**Search Strategy**

This is the third and fourth step where search strings are defined and search is conducted in different levels. The search strategy for SLR was primarily adopted from the report of Kitchenham and Charters [13]. The main concern of this process is the creation of a search string. The primary search was done using two simple keywords, "sustainability" and "requirements" in the databases listed above. The Boolean operators "AND" and "OR" are applied to broaden searches by including synonyms, alternative spellings, or closely related terms. The results of the primary search were focused on identifying SRs mapped for software or information systems. However, most of these works were refined to only the environmental dimension. The search was extended to next level by including terms related to different dimensions of sustainability like "Social", "Economic". As a result, we obtained research articles that primarily focused on specific dimensions of sustainability for software and information systems. Another level of search was done considering specific application domains such as smart systems, cloud computing, and healthcare. This search was necessary as there as many application-specific works focusing on sustainability requirements, but they are often indexed later when searched using generic keywords. However, the search is restricted only to specific domains. As the survey part of the research work was time-bounded.

Table 1 provides a detailed record of the search strings and the number of records obtained. The first row of table 1 contains the search string used for the initial primary search. The following three rows contain the search string used for second level of search based on sustainability dimensions. Finally to search for different domains the search strings in the first four rows has to be concatenated with a specific domain name. An example of a third-level search string is given in the last row of the table.

The search strategy employed using keywords in Table 1 helps to answer the research question *RQ-1*. These also helped to identify the categories within each dimension and also metrics associated with each category. The metrics help to assess whether SRs are satisfied for a particular software system. However, not every research article mentioned a proper metric. We have separately applied a similar search strategy in the digital databases to identify metrics and measures for different types of sustainability concerns within each dimension (refer to figure **??**). The keywords that are additionally used are- "Sustainability measure" OR "sustainability metric" or "sustainability measurement".

**Inclusion and Exclusion Criteria**

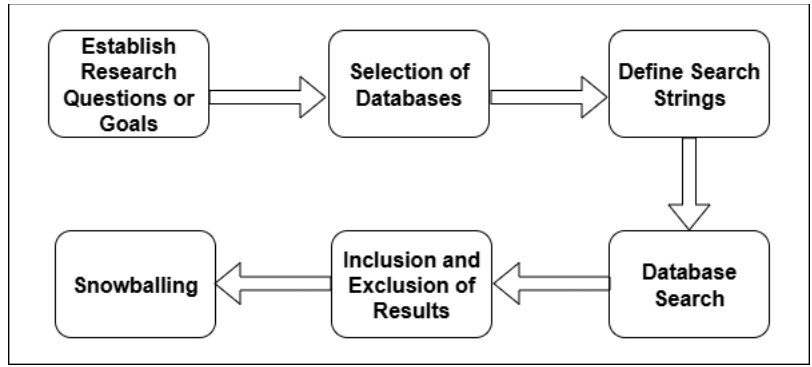

Figure 1: Overview of SLR

Table 1: Database Search Strings

| Search String | Number of Records in each Database Found | | | |
| --- | --- | --- | --- | --- |
| | Scopus | IEEE Xplore Digital Library | ACM Digital Library | Wiley Online Library |
| "sustainability requirement" OR "sustainable requirement" OR "sustainability principle" OR "sustainable principle" | 2863 | 18 | 15822 | 110 |
| "social sustainability requirement" OR "social sustainable requirement" OR "social sustainability principle" OR "social sustainable principle" | 886 | 1850 | 22,285 | 4735 |
| "technical sustainability requirement" OR "technical sustainable requirement" OR "technical sustainability principle" OR "technical sustainable principle" | 170 | 2241 | 33,088 | 4735 |
| "economic sustainability requirement" OR "economic sustainable requirement" OR "economic sustainability principle" OR "economic sustainable principle" | 6321 | 2897 | 14807 | 0 |
| ("sustainability requirement" OR "sustainable requirement" OR "sustainability principle" OR "sustainable principle) AND ("Healthcare") | 8059 | 368 | 7071 | 2593 |

In this step, criteria are established for the selection of works from the search results. The inclusion and exclusion criteria ensure that only studies directly relevant to the research questions are selected. The inclusion criteria applied in this research for constructing the taxonomy are as follows-

(1) The top 100 records found for each search query are considered.
(2) Peer-reviewed research works published in journals, conferences and workshops.
(3) Research works that are written in English.
(4) Research works from the domain of computer science or software engineering.
(5) Studies where sustainability requirements for any software system are explicitly mentioned in the title and abstract.
(6) Studies that explicitly mention sustainability frameworks, principles or catalogues. These studies help in particular, to identify categories and metrics.

The exclusion criteria are as follows-

(1) Studies published before 2010.
(2) Studies that discussed sustainability broadly but did not explicitly address requirements, their classification, or detailed definitions.
(3) Non peer-reviewed studies like tutorials, posters and other such.

The studies that satisfied the inclusion and exclusion criteria are selected from the initial search. The total number of articles obtained from initial search was $x$ and out of which $y$ was selected for constructing the taxonomy. The research articles that are common across different databases are considered only once during the initial search results.

**Snowballing**
In the last step, snowballing is applied to consider the reference list in the selected research works. The citations made to the selected research works are also considered. The inclusion and exclusion criteria are also applied to these referenced research works. The snowballing is done only once based on the works selected in the previous step.

### 3.3 The Taxonomy
Each of the selected papers was carefully analyzed to extract the SRs, their categories and metrics (if mentioned). This process is done manually by the authors and also taking the help of AI tools for knowledge extraction. The SRs mentioned in each research work are at first listed along with its dimension and category.

Table 2 shows an abridged version of the sustainability requirements taxonomy. The taxonomy captures lists of sustainability requirements that must be satisfied for any software system. Let us consider the SR "Changes can be incorporated effectively"- this is also a general quality requirement of software. However, it is an important SR as it ensures the products success or failure and directly impacts the economic sustainability. Another SR- "Appropriate tones should be chosen in a color gradient that depends on the particular technology of the monitor"- is an important social sustainability concern as it has direct emotional impact and also cultural sensitivity. Whether these SRs are relevant for a particular software system is determined by the *Requirement Analysis Framework* (described in the following section) based on the particular product scope.

Table 2: Sustainability Requirements Taxonomy (abriged)

| Sustainability Requirement | Dimension | Category |
|---|---|---|
| Assessing software over a long term and consider decisions with this in mind. | Technical | Longevitivity |
| Identify the most effective way to be sustainable over alternative options at different levels to ensure the most sustainable choice is being made. | Social | Equitable Access and Inclusivity |
| System must achieve pragmatic goals | Social | Inclusitivity |
| System must ensure confidence in the organization | Social | Social Engineering |
| System components can be separated and recombined | Technical | Repairability |
| Changes can be incorporated effectively | Technical | Perdurability |
| An appropriate low-energy color scheme should be chosen based on the monitor technology | Environmental | Energy Consumption |
| Appropriate tones should be chosen in a color gradient that depends on the particular technology of the monitor | Social | Social Engineering |
| The number of progress bars, animations, and especially scrollbars used, should be reduced. | Technical | Feature Bloat |
| Menus should present as few choices as possible to reduce the human cognitive process of decision-making, in line with Hick-Hyman Law | Technical | Feature Bloat |
| Information shown in the GUI should be divided with thematic icons to reduce interaction complexity and allow for quick information retrieval | Social | Social Engineering |
| Excessive use of computer graphics processing units should be avoided | Environmental | Energy Consumption |
| Low energy-consuming states of graphics processing units should be used | Environmental | Energy Consumption |
| Screen changes during the performance of tasks should be reduced | Environmental | Energy Consumption |
| Autocomplete functionality with at least three initial letters should be developed to quickly write complex names | Social | Equitable Access |
| Systems must be interpretable and adaptable to changing goals, dynamic data, unexpected failures, and security threats | Technical | Longeitivity |
| Every in-efficient code patterns should be resolved | Environmental | Carbon Footprint |
| Minimizing unnecessary cognitive load on users. | Social | Social Engineering |
| Ensuring software systems do not perpetuate or exacerbate social inequalities. | Social | Equitable Access |
| Ensuring systems promote fair market practices. | Economic | Affordability |

The taxonomy is exhaustive and non-repetitive and it also contains requirements that may pertain to specific categories of systems (derived from research works of specific domains). The taxonomy is open-sourced [1] and the list of SRs may be updated periodically. A dataset separately for categories and metrics of each dimension are also created. Table 3 and 4 provide the overview of categories and metrics that are found in this research work through SLR. Some of these metrics can be measured during the development phase, while other can only be measured based on the deployed software. For example, the category customer social value within social dimension can be measured using metric amount of positive feedback. Such a metric value can only be obtained when the system is used. There are some categories for which multiple metrics could be found. Energy consumption category within environmental dimension consists of multiple metrics. In case of some metrics, a well-defined formula exists such as software carbon intensity, development cost, degree of cohesion and coupling and others. But for some metrics no well-defined formulas are there such as amount of non-reusable modules, well-being and many more. The value of those metrics can be interpreted based on the judgment of a system analyst. The list of state-of-the-art works referred for building the taxonomy, categories, metrics and measures are provided in the GitHub repository [1].

## 4 Correlation among Sustainability categories

Each sustainability dimension encompasses various categories as shown in Tables 3 and 4. These categories influence each other. For example E-waste category of *Environmental* dimension may correlate with the circular economy category of *Economic* dimension. Reusing and recycling software components leads to less wastage. Similarly, feature bloat in the *Technical* dimension may have a negative correlation with energy consumption, as it leads to unnecessary processing power and increased emissions. These correlations are not explicitly define. In this work, we apply an large language model (LLM) based approach to construct a catalog of sustainability category correlations. The steps are as follows-

(1) A dataset is built, consisting of different categories of each dimension and their respective definitions.
(2) This dataset is prompted to LLM to generate a correlation matrix.
(3) The generated correlation matrix are then presented to each author separately. Each authors marks whether they agree or not with the defined correlations.
(4) The results obtained from each author are combined in a single worksheet.

(5) Finally, the combined worksheet and presented for a group discussion among the authors. The final correlations are discussed among all the authors to decide which correlations are reasoning are appropriate. The final correlation matrix is based upon the knowledge and voting of all the authors.

In the generation process, we have used LLMs GPT3.5 [23] and Gemini 2.5 Pro [12]. We found both the LLMs have generated similar results and reasoning. Tables 5 and 6 shows the positive and negative correlations among LLM categories, respectively. These correlation matrix results are not necessarily absolute; that is, the relation may hold based on context or scenario. Hence, whether a particular correlation should be considered in the design depends upon the product context or requirements.

## 5 Conclusion

This research work provides two main contributions. The first contribution consists of building a general taxonomy of sustainability requirements obtained from various existing research works. Software developers may use this taxonomy to find specific sustainability concerns for a particular software domain. The taxonomy is open-sourced and can be iteratively populated with more data. The second contribution is the correlation matrix among different sustainability categories. These correlations define how across the dimensions different issues may have a positive or negative impact on each other. This is important as software developers need to maintain a trade-off among all concerned categories while building the product, to make it sustainable.

## Acknowledgement

Work partially funded by the EcoDigify Erasmus+ project co-funded by the European Union (Project Reference: 2024-1-SE01-KA220-HED-000250071).

## Bibliography

[1] Maryam Al Hinai and Ruzanna Chitchyan. 2014. Social sustainability indicators for software: Initial review. *CEUR Workshop Proceedings* 1216 (01 2014), 21–27.
[2] Ahmed D. Alharthi, Maria Spichkova, and Margaret Hamilton. 2019. Sustainability requirements for eLearning systems: a systematic literature review and analysis. *Requir. Eng.* 24, 4 (Dec. 2019), 523–543. doi:10.1007/s00766-018-0299-9
[3] Ahmed D. Alharthi, Maria Spichkova, Margaret Hamilton, and Tawfeeq Alsanoosy. 2018. Gender-Based Perspectives of eLearning Systems: An Empirical Study of Social Sustainability. In *Integrated Spatial Databases*. https://api.semanticscholar.org/CorpusID:150338679
[4] Christoph Becker, Stefanie Betz, Ruzanna Chitchyan, Leticia Duboc, Steve M. Easterbrook, Birgit Penzenstadler, Norbet Seyff, and Colin C. Venters. 2016. Requirements: The Key to Sustainability. *IEEE Software* 33, 1 (2016), 56–65. doi:10.1109/MS.2015.158
[5] Christoph Becker, Ruzanna Chitchyan, Leticia Duboc, Steve Easterbrook, Birgit Penzenstadler, Norbert Seyff, and Colin C. Venters. 2015. Sustainability Design and Software: The Karlskrona Manifesto. In *2015 IEEE/ACM 37th IEEE International Conference on Software Engineering*, Vol. 2. 467–476. doi:10.1109/ICSE.2015.179
[6] Ruzanna Chitchyan, Christoph Becker, Stefanie Betz, Leticia Duboc, Birgit Penzenstadler, Norbert Seyff, and Colin C. Venters. 2016. Sustainability design in requirements engineering: state of practice. In *Proceedings of the 38th International Conference on Software Engineering Companion* (Austin, Texas) *(ICSE '16)*. Association for Computing Machinery, New York, NY, USA, 533–542. doi:10.1145/2889160.2889217
[7] Claudia-Melania Chituc. 2023. On the Importance of a Requirements Elicitation Framework to Ensure Sustainability in the Digital Education Ecosystems. In *2023 IEEE 31st International Requirements Engineering Conference Workshops (REW)*. 352–356. doi:10.1109/REW57809.2023.00066

---

[1] https://github.com/MCompRETools/SustainabilityTaxonomy

Table 3: Categories of Environmental and Technical Dimension

| Dimension | Category | Metric | Measure |
|---|---|---|---|
| Environmental | Carbon Footprint | Software Carbon Intensity | Numeric |
| | Energy Consumption | Energy efficiency, Runtime efficiency, CPU-intensity, memory usage, peripheral intensity, idleness and Algorithmic efficiency | |
| | E-waste | Amount of non-reusable modules | |
| | Code Sustainability | CPU usage, Memory usage, Code smells | Percentage |
| | Resource management | Energy efficient data management, Water footprint | Numeric, Percentage |
| Technical | Perdurability | Technical evolution, functional evolution | Numeric- change frequency, function-point analysis |
| | Longitivity | Mean time between failures, Aveage life span in usage hours per year | Numeric |
| | Repairability | Degree of Cohesion and coupling, Degree of traceability docuementation | |
| | Feature Bloat | Percentage of user engagement with each feature | |
| | Security Concerns | Number of known vulnerabilities | |

Table 4: Categories of Social and Economic Dimension

| Dimension | Category | Metric | Measure |
|---|---|---|---|
| Social | Digital Inclusivity | Number of supported language, demographics | Numeric |
| | Equitable Access | End-user data availability, horizontal and vertical equity | Qualititative |
| | Ethical Concerns | Degree of bias | |
| | Customer social value | Amount of positive feedbacks | Percentage |
| | Social Engineering | Well Being | Qualitative |
| | User error protection | Recovery Time | Numeric |
| Economic | Circular Economy | Percentage of reused code or services, Amount of shared infrastructure | Percentage |
| | Affordability | Access Cost | Numeric |
| | Cost efficiency | Development cost- number of workers, amount of time and effort | |
| | Software process evolving intellectual capital | Customer capital value, Market requirements value, innovation value, maintenance value | Qualitative and numeric |

[8] Nelly Condori-Fernandez and Patricia Lago. 2018. Characterizing the contribution of quality requirements to software sustainability. *Journal of Systems and Software* 137 (2018), 289–305. doi:10.1016/j.jss.2017.12.005

[9] Theresia Ratih Dewi Saputri and Seok-Won Lee. 2021. Software sustainability requirements: a unified method for improving requirements process for software development. In *2021 IEEE 29th International Requirements Engineering Conference (RE)*. 506–507. doi:10.1109/RE51729.2021.00077

[10] Matthew J Eckelman, Ulli Weisz, Peter-Paul Pichler, Jodi D Sherman, and Helga Weisz. 2024. Guiding principles for the next generation of health-care sustainability metrics. *The Lancet Planetary Health* 8, 8 (2024), e603–e609. doi:10.1016/S2542-5196(24)00159-1

[11] José A. García-Berná, José L. Fernández-Alemán, Juan M. Carrillo de Gea, Ambrosio Toval, Javier Mancebo, Coral Calero, and Félix García. 2021. Energy efficiency in software: A case study on sustainability in personal health records. *Journal of Cleaner Production* 282 (2021), 124262. doi:10.1016/j.jclepro.2020.124262

[12] Eric Bieber et al. Gheorghe Comanici. 2025. Gemini 2.5: Pushing the Frontier with Advanced Reasoning, Multimodality, Long Context, and Next Generation Agentic Capabilities. *arXiv* (2025). https://arxiv.org/abs/2507.06261

[13] Barbara Kitchenham and Stuart Charters. 2007. Guidelines for performing systematic literature reviews in software engineering technical report. *Software Engineering Group, EBSE Technical Report, Keele University and Department of Computer Science University of Durham* 2 (2007).

[14] Christoph König, Daniel J. Lang, and Ina Schaefer. 2025. Sustainable Software Engineering: Concepts, Challenges, and Vision. 34, 5, Article 135 (May 2025), 28 pages. doi:10.1145/3709352

[15] Hira Noman, Naeem Ahmed Mahoto, Sania Bhatti, Hamad Ali Abosaq, Mana Saleh Al Reshan, and Asadullah Shaikh. 2022. An Exploratory Study of Software Sustainability at Early Stages of Software Development. *Sustainability* 14, 14 (2022). doi:10.3390/su14148596

**Table 5: Positive Correlations among Sustainability categories**

| Category 1 | Category 2 | Reason |
|---|---|---|
| Carbon Footprint ↓ | Energy Consumption ↓ | Lower energy use reduces GHG emissions. |
| Energy Consumption ↓ | Code Sustainability ↑ | Optimized code reduces computing power and energy demand. |
| E-waste ↓ | Circular Economy ↑ | Reuse and recycling reduce waste volumes and toxicity. |
| Code Sustainability ↑ | Longevity ↑ | Efficient, maintainable code stays viable longer. |
| Longevity ↑ | Repairability ↑ | Easily repairable systems remain in use longer. |
| Repairability ↑ | Circular Economy ↑ | Repair supports reuse and recycling loops. |
| Resource Efficiency ↑ | Cost Efficiency ↑ | Minimizing waste reduces input costs. |
| Perdurability ↑ | Software Process Evolution ↑ | Long-lived systems benefit from adaptive processes. |
| Digital Inclusivity ↑ | Equitable Access ↑ | Removing access barriers helps more people use digital systems. |
| Digital Inclusivity ↑ | Customer Social Value ↑ | Wider access increases community benefit. |
| Ethical Concerns (well-handled) ↑ | Customer Social Value ↑ | Responsible practices increase trust and social impact. |
| Affordability ↑ | Equitable Access ↑ | Lower costs make technology accessible to more users. |
| Circular Economy ↑ | Affordability ↑ | Reuse/recycling can lower costs over time. |
| Software Process Evolution ↑ | Code Sustainability ↑ | Process improvement leads to maintainable code. |

**Table 6: Negative Correlations among Sustainability categories**

| Category 1 | Category 2 | Reason |
|---|---|---|
| Feature Bloat ↑ | Energy Consumption ↑ | More features increase resource use and energy demand. |
| Feature Bloat ↑ | Carbon Footprint ↑ | More processing increases emissions. |
| Feature Bloat ↑ | Longevity ↓ | Overloaded systems become obsolete faster. |
| Feature Bloat ↑ | Repairability ↓ | Complex designs are harder to fix. |
| Repairability ↑ | Cost Efficiency ↓ (short-term) | Repair-friendly design can increase initial costs. |
| Longevity ↑ | Software Process Evolution ↑ (potential tension) | Long-lived systems may resist change. |
| Security Concerns ↑ | Affordability ↓ | Stronger security can increase costs. |
| Ethical Concerns ↑ | Cost Efficiency ↓ (short-term) | Ethical compliance may require costly processes. |
| Digital Inclusivity ↑ | Cost Efficiency ↓ | Inclusivity may increase development costs. |
| Affordability ↑ | Carbon Footprint ↑ (if poorly managed) | Cheap production can be unsustainable. |
| Resource Efficiency ↑ | Repairability ↓ (possible) | Highly integrated designs can be harder to repair. |
| Circular Economy ↑ | Cost Efficiency ↓ (short-term) | Recycling may require costly infrastructure. |
| Social Engineering ↑ | Security Concerns ↑ | Manipulation risks undermine security. |
| User Error Protection ↑ | Cost Efficiency ↓ | Designing error-tolerant systems can require extra resources. |

[16] Sofia Ouhbi, José Fernández-Alemán, Ambrosio Toval, José Rivera, and Ali Idri. 2017. Sustainability requirements for connected health applications. *Journal of Software: Evolution and Process* 30 (11 2017), e1922. doi:10.1002/smr.1922

[17] Sofia Ouhbi, Ali Idri, and José Luis Fernández-Alemán. 2018. *Standards-Based Sustainability Requirements for Healthcare Services in Smart Cities*. Springer International Publishing, Cham, 299–317. doi:10.1007/978-3-319-76669-0_13

[18] Yen Dieu Pham, Abir Bouraffa, Marleen Hillen, and Walid Maalej. 2021. The Role of Linguistic Relativity on the Identification of Sustainability Requirements: An Empirical Study. In *2021 IEEE 29th International Requirements Engineering Conference (RE)*. 117–127. doi:10.1109/RE51729.2021.00018

[19] Yen Dieu Pham, Abir Bouraffa, and Walid Maalej. 2020. ShapeRE: Towards a Multi-Dimensional Representation for Requirements of Sustainable Software. In *2020 IEEE 28th International Requirements Engineering Conference (RE)*. 358–363. doi:10.1109/RE48521.2020.00048

[20] Theresia Ratih Dewi Saputri and Seok-Won Lee. 2020. Addressing sustainability in the requirements engineering process: From elicitation to functional decomposition. *J. Softw. Evol. Process* 32, 8 (Aug. 2020), 25 pages. doi:10.1002/smr.2254

[21] Siraj Shikhli, Amir Shikhli, Anwar Jarndal, Imad Alsyouf, and Ali Cheaitou. 2022. Towards Sustainability in Buildings: a Case Study on the Impacts of Smart Home Automation Systems. In *2022 Advances in Science and Engineering Technology International Conferences (ASET)*. 1–8. doi:10.1109/ASET53988.2022.9735126

[22] Angelos Stoumpos and Michael Talias. 2021. *Economic Sustainable Health Information Systems*. 234–251. doi:10.4018/978-1-7998-5442-5.ch012

[23] Junjie Ye, Xuanting Chen, Nuo Xu, Can Zu, Zekai Shao, Shichun Liu, Yuhan Cui, Zeyang Zhou, Chao Gong, Yang Shen, Jie Zhou, Siming Chen, Tao Gui, Qi Zhang, and Xuanjing Huang. 2023. A Comprehensive Capability Analysis of GPT-3 and GPT-3.5 Series Models. *arXiv* (2023). https://arxiv.org/abs/2303.10420